# Teardown Analysis of Samsung S20 Exynos 990 SoC

Nabeel Ahmad Khan Jadoon
101410832
nabeelahmadkhan.jadoon@aalto.fi

Daryna Ihnatiuk
101073651
daryna.ihnatiuk@aalto.fi

Umama Saleem
101818971
umama.saleem@aalto.fi

Paulina Gallego Pérez
101385653
paulina.gallegopere@aalto.fi

Joanna Ylänen
44912T
joanna.ylanen@aalto.fi

*Abstract*— The mobile phone has evolved from a simple communication device to a complex and highly integrated system with heterogeneous devices, thanks to the rapid technological developments in the semiconductor industry. Understanding the new technology is indeed a time-consuming and challenging task. Therefore, this study performs a teardown analysis of the Samsung Exynos S20 990 System-on-Chip (SoC), a flagship mobile processor that features a three-dimensional (3D) package-on-package (PoP) solution with flip chip interconnect (fcPoP). The fcPoP design integrates the SoC and the memory devices in a single package, reducing the interconnection length and improving signal integrity and power efficiency. The study reveals the complex integration of various components and the advanced features of the SoC. The study also examines the microstructure of the chip and the package using X-ray, SEM, and optical microscopy techniques. Moreover, it demonstrates how the fcPoP design enables the SoC to meet the demands of higher performance, higher bandwidth, lower power consumption, and smaller form factor, especially in 5G mobile applications. The study contributes to understanding advanced packaging methodologies and indicates potential directions for future semiconductor innovations.

## I. Introduction

Samsung Galaxy S20 is one of the flagship smartphone series released in 2020 featuring a powerful system-on-chip (SoC) called Exynos 990. It is a mobile high-end SoC capable of integrating three clusters of processor cores with different architectures of graphics processing unit (GPU), a neural processing unit (NPU), and cellular modems. It is fabricated using 7nm process technology and supports 16 GB of LPDDR5 memory. LPDDR5, which stands for Low Power Double Data Rate 5, offers dynamic random-access memory (DRAM) providing high performance and low power consumption for mobile devices [1]. The chip's memory is divided into two memory stacks to reduce the wire bonding length and to enhance the signal integrity between the memory and the logic components [2]. The signals are accessed through the Au wire bonding technique used to create the interconnections between the memory components and SoC.

Research shows that smaller process nodes tend to have a shorter lifetime due to higher transistor density and lower voltage margins [3]. Therefore, these kinds of chips are prone to aging effects such as electromigration, hot carrier injection, and bias temperature instability. The power consumption of the Exynos 990 can also affect its lifetime as different scenarios impose different levels of stress at consumer-level applications. Even though Samsung has implemented the distributed heat transfer technology on this SoC, the lifetime for this chip is roughly estimated to be around 3 to 5 years in normal usage. However, this can be further reduced by user applications such as video streaming, web browsing, and messaging [4]. The Exynos 990 is claimed to offer intelligent capabilities, improved performance, and power efficiency but it has also faced criticism as compared to its counterpart Qualcomm Snapdragon 865 SoC which powers the Chinese variants of Galaxy S20 smartphones [5].

The typical architecture of this SoC is Package-on-Package (PoP) technology, which stacks two packages on top of each other using FCBGA and attaches them to the motherboard using a single interface. This PoP technology enables higher integration. The components are heterogeneously connected in 3D stacking and electrically connected using various interconnection technologies like Wire bonding, Flip Chip Ball Grid Array (FCBGA), and Re-Distribution Layers (RDLs).

Wire bonding in this SoC is a relatively simple, cost-efficient technique applied to the interconnection of individual ICs and chips to the package [6]. It is the most common and conventional contact-forming approach used in device packaging. Typically, contacts for electrical signal propagation are made of Au, Cu, or Al wires bonded to the contact pads of 2 devices being connected. Contact pads are located on the outer edges of the dies to reduce the connection length as much as possible. Wiring is done through thermos-compressive, thermosonic, or ultrasonic assistance with the formation of ball or wedge bonds depending on the process and material requirements [7]. Despite its advantages and applicability, this contact forming method has shown multiple drawbacks and limitations, like area consumption due to I/O pads positioning on the chip surface, signal distortion at high frequencies due to the length of connections, or electrical and mechanical contact degradation due to aging of the contact interface.

On the other hand, the FCBGA (flip chip ball grid array) is a high-performance semiconductor packaging solution that utilizes compact and efficient technology as compared to wire bonding. FCBGA provides design flexibility for higher signal density and functionality in a smaller die and packaging footprint. For the flip chip connections either Control Collapse Chip Connection (C4) or Chip Connection (C2) process can be used and has been discussed later. It is often used in ASICs, DSPs, and various other high-performance applications, including products such as Samsung Microprocessors, Intel's Core i7, Texas Instruments custom ASICs, and Freescale processors [8][9]. FCBGA offers several benefits like enhanced signal integrity, reduced inductance, and improved transmission of signals. It also provides mechanical reinforcement and load sharing, reducing strain on the solder joints during temperature cycling from assembly to board attachment and operation. However, it is important to note that FCBGA is not an affordable packaging method and is typically used in applications where performance is more critical than cost [10].

For flip chip connections, alternative wafer bumping methods are used such as C4 and C2. The difference between the processes is the solder ball collapses in C4 connections during the reflow soldering, which doesn't happen in C2, since the Cu bumps don't melt. therefore, the final connection height is closer to the original, un-soldered bump height than with C4 bumps. C2 bumps also have higher thermal conductivity, lower electrical resistivity, and capability for finer pitch compared to C4, which can go down to 50 µm in C2 type [8], but the downsides are poor self-alignment and the cost. However, C4 connections are relative to BGA solder connections technology that is well known and used for decades.

For all the interconnections discussed earlier, intermetallic mixing is very important. It is a phenomenon that can occur when metals have immediate contact with each other. Under certain circumstances, metals tend to diffuse one into another and form intermetallic compounds. E.g., For exploring the reliability perspective, the contact between gold and aluminum can cause the formation of intermetallic compounds (IMC) like $Au_5Al_2$ (white plague) and $AuAl_2$ (purple plague) [6]. These IMCs can increase the electrical resistance, create voids and cracks in metal near the interface, and reduce the bond strength. It is recommended to avoid the high temperatures and ultrasonic parameters for direct contact of these bonds. Thus, the formation of different compositions in contact areas needs to be tracked and requires more thorough research and evaluation of possible negative effects on device performance over a lifetime.

The rapid development in the semiconductor field offered advanced packaging technologies and the complex knowledge gap behind it. It is critical to know the intricate microarchitecture and interconnections to effectively evaluate the inherent complexities in the processes. To understand the various facets of this SoC, this paper aims to provide a comprehensive overview of the packaging technology behind the Exynos 990 SoC. Our objective is to unlock the potential to understand different components' architecture and seek to contribute crucial insights into advanced packaging like PoP. For this work, a teardown study at Aalto University is made to identify the package structure, interconnection technologies, and materials used for this SoC.

The methods used included cross-section analysis with optical microscopy and scanning electron microscopy (SEM) as well as X-ray analysis.

Optical microscopy is one of the most basic, yet informative methods for sample characterization and analysis. It allows the magnification of different sample features up to 100x times by different optical lenses. This method has provided extensive visual information on the cross-section of the device.

The main principle of SEM is based on the interaction of the high-energy electron beam with the sample surface. Emitted electrons cause the scattering of electrons from the surface of the sample and either get backscattered or diffracted backscattered or cause secondary electrons emission. All scattered electrons are captured for further conversion into the image with a high depth of field. To avoid charging the sample surface, the sample was covered with 10 nm of Cr film beforehand. SEM imaging was performed with JEOL JSM-7500FA (Analytical high-resolution SEM) at 15 keV with secondary and backscattered electrons measured. EDX imaging was performed with the same tool, but a different detector was used for that purpose.

X-ray imaging was performed with Procon X-Ray Fraunhofer at different angles (cross-, top and angled imaging). This method utilizes a higher energy beam of 50 keV and allows the detection of transmitted electrons through the sample allowing the reconstruction of imaged sample features. Different materials belonging to the sample detain electrons with different efficiency, resulting in contrast differences in the reconstructed image.

The results obtained from imaging sessions were supported by the binary phase diagram analysis of the corresponding materials. Our investigation highlighted the importance of fcPoP structure in optimizing signal integrity for consumer applications in mobile devices. These insights not only help to understand the microarchitecture of the SoC but also pave the way for future semiconductor innovation considering critical parameters like power, form factor, and chip failures.

The following Results section will present the findings; first on the device level, then concentrating on the separate parts defined by the device. The discussion section will analyze the findings and their meanings. In the end, the Conclusion summarizes the study and highlights the key findings.

## II. RESULTS

### II.I DEVICE LEVEL

The device under analysis was a PoP component presented in the stitched SEM image in Fig. 1. The upper package consists of 8 memory dies and a processor die (Fig.2) and buried capacitor(s) (Fig.3) in the lower package. The memory dies are stacked with the die attach adhesive with bond line thicknesses (BLTs) 20…40um (Fig.4). Both packages use over-molding technology with the BGA connections, but the molding compounds look different compared to each other; larger fillers in the memory package's epoxy mold compound (EMC). The interposers used in the packages are organic: the memory package has a 3-layer interposer, and the processor package has a 4-layer interposer under the die and a 2-layer interposer for stacking the memory package on top of it. The interposers are organic.

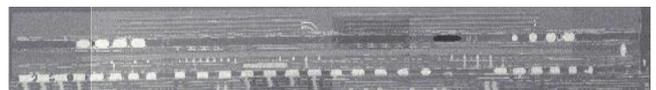

*Figure 1.* Stitched SEM image of the PoP component structure of Samsung Exynos 990.

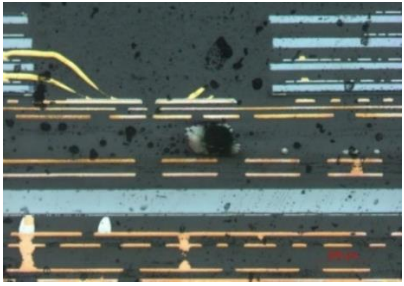

*Figure 2.* Optical microscopy image (5X) of the dies in the PoP package: 2 stacks of 4 memory die wire bonded onto the upper package's interposer, top substrate of the lower package, processor die with C2 interconnections to the interposer of the lower package.

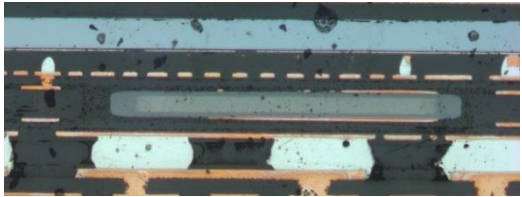

*Figure 3.* Optical microscopy image (5X) of a buried capacitor in the bottom substrate of the processor package.

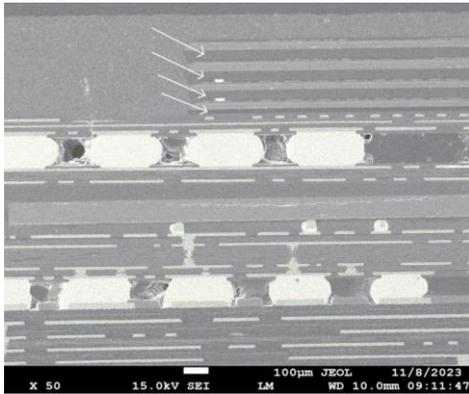

*Figure 4.* SEM image pointing out the die attach adhesive layers in the memory stacks.

The C4/BGA connections were used in three locations in the package (Fig.5): between the memory package and the processor package, between the processor's interposer and the connecting interposer, and for the device-level connections to the PCB. The EDS analysis confirmed the structure of all solder connections being the same: Cu pads on the interposers covered with Ni/Au coating, probably electroless Nickel Immersion Gold (ENIG), and the solder used in all being the same lead-free solder, SnAgCu (SAC) as some Cu can be seen all over the EDS mapped area of one of the solder joints (Fig.6). The closer analysis of the solder connections also showed intermetallic layer formation on the interfaces of the Cu pad and the solder (Fig.7). The only deviating detail was the solder mask design. On the interposers, all pads are solder mask defined (SMD) but the PCB pad design is non-solder mask defined (NSMD) as seen e.g., in Fig. 5.

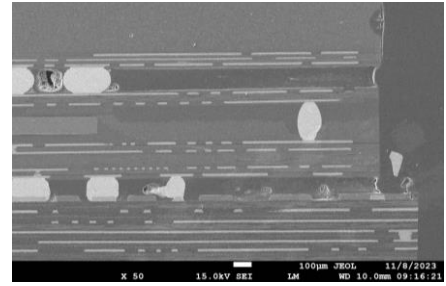

*Figure 5.* SEM image presenting the C4/BGA connections between the memory package and the processor package, between the interposers of the processor package, and between the device and the PCB.

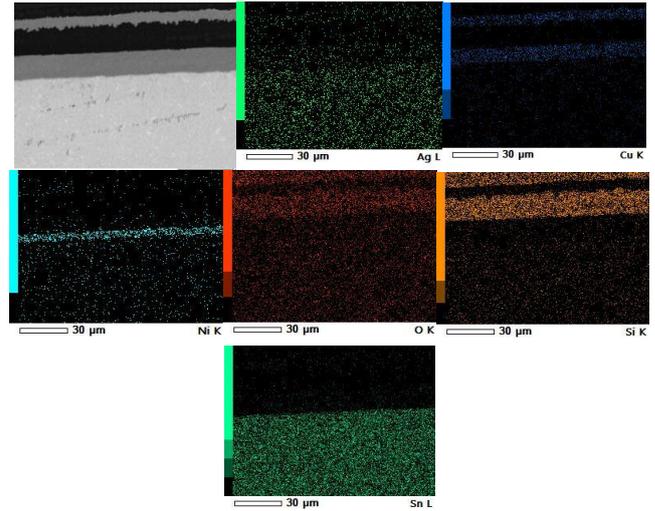

*Figure 6.* SEM image of solder joint area with respective EDS mapping of elements.

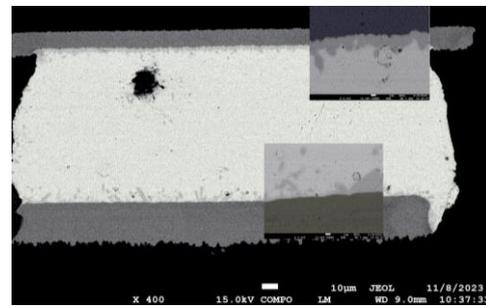

*Figure 7.* SEM images of one of the BGA solder joints. A large void is seen on the left and intermetallic layers growing on the interfaces of the pad and solder.

The analysis showed that there are several I/O layouts in the device, even among the same connection layer. The memory package has BGA connections in four rings as seen e.g., in Fig.1. The ball size is approximately 250um, the ball height is 125um, and the pitch is 400um. The C4 connections between the processor interposers are positioned mainly in a single ring with two additional rows at the package ends (Fig.1 and 8). The ball's diameter is approximately 75um, the ball's height is 190um, and the pitch is 125um. The layout of the BGA connections from the device to the PCB is the most complex as seen in Fig.8. There is a matrix of 14 connections in a row in the middle area with approximate ball size 250um, ball height 80um and 470um pitch in both x and y axis directions. Around this matrix are the main connections with the same ball size and height. However, the approximate pitch in the x-axis direction (cross section's surface) is 400um on every row but in the y-axis direction the solder balls form a zigzag pattern as on every second row/column the balls are in line to each other while on the rows/columns in between the balls are positioned in the middle of the previous line's balls.

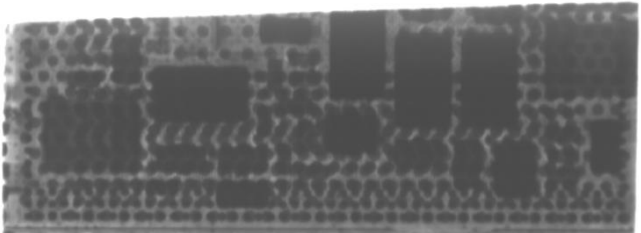

*Figure 8. X-ray image of the device's connection layout.*

## II.II MEMORY PACKAGE

The memory component in Samsung Exynos 990 SoC is stacked on four layers that are divided into right and left half shown in Fig. 9. Communication between memory and processor components is enabled through wire bond interconnects.

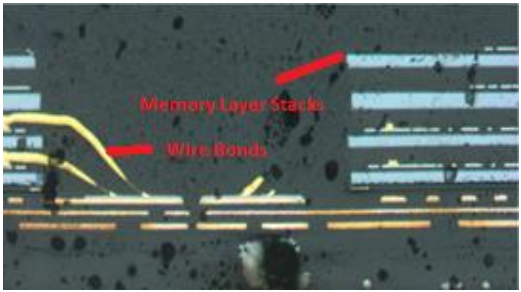

*Figure 9. Optical Microscopy of Memory Stacking*

The EDS mapping of the wire bonds (Fig. 10 and 11) helped to identify elements that comprise the contact areas and ratio between elements in the area of intermetallic mixing of Au-Al contact. While Au wire relates to a "ball" bond to the Al pad with a thin Ti barrier layer underneath on the memory side (Fig. 9), on the processor side it is attached with a "wedge" bond to the contact area with a stack of Au/Ni/Cu (Fig.11).

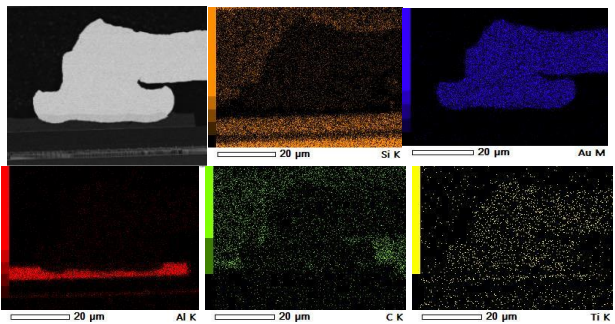

*Figure 10. SEM image of "ball" bond area with respective EDS mapping of elements.*

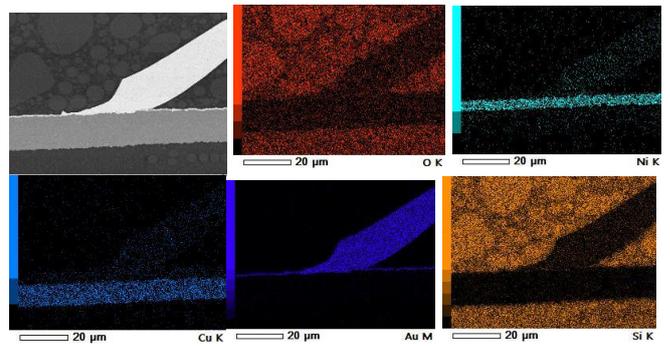

*Figure 11. SEM image of "wedge" wire bond area with respective EDS mapping of elements.*

The data in Table 1 shows that the material has a high mass percentage of Au (around 96%) and a low quantity of Al (around 3%) at all the points marked in Fig 10. The atomic percentage of Au and Al is 80% and 20% respectively. Estimated thicknesses for the respective contact layers would be approx. 20 µm for Au wire itself, 5 µm for Al pad with 350-400 nm layer of Ti underneath on the memory side, and 1 um Au/ 5-6 um Ni/ 10 µm Cu stack on the other side of the wire attachment.

*Table 1. EDS analysis results for Au-Al mass composition*

| Point | Element | Atom% | Mass% |
|---|---|---|---|
| 2 | Al K | 19.82 | 4.0977 |
| | Au M | 80.18 | 95.9023 |
| 3 | Al K | 20.08 | 4.1651 |
| | Au M | 79.92 | 95.8349 |

## II.III PROCESSOR PACKAGE

The Exynos 990 processor package employs a flip-chip ball grid array (FCBGA) packaging technique improving electrical and thermal characteristics. This involves solder bumps connecting the active sides of the chips to the interposer. The processor die has two interposers, the upper with 2 Cu layers and the lower with 4 Cu layers as seen in Fig. 12.

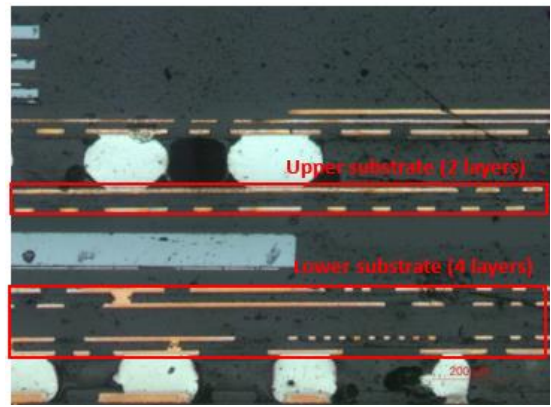

*Figure 12. Processor Package with upper and lower substrate layers.*

The processor package makes use of five different interconnections. It utilizes two FCBGA arrays to connect with both the memory package and the PCB, solder balls to connect the upper and lower substrates of the processor package itself as shown in Fig. 13, Cu-based interconnections in the substrate RDL and C2 type flip chip connections to connect the processor to the substrate as shown in Fig. 14.

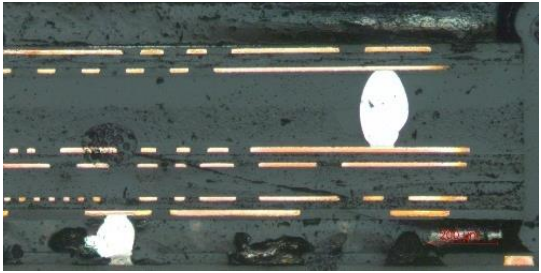

*Figure 13. Solder connection between upper and lower processor package substrates.*

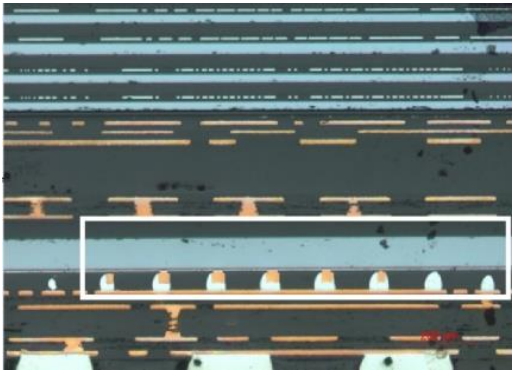

*Figure 14. C2 bumps connect the processor to the interposer.*

On the active side of the processor are numerous tiny Cu pillar micro-bumps with Sn-solder caps, approximately 54 µm in diameter. For denser connections, the processor chip uses micro-bumps at a pitch/spacing of 154 µm with the bump height being 61 µm approximately as shown in Fig. 15. A smaller bump pitch allows for massive I/O counts [9]. It is also seen that the interposer is organic containing Cu layers with epoxy resin and glass fibres.

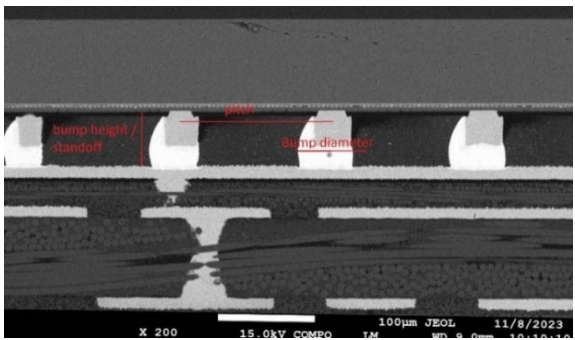

*Figure 15. SEM image of C2 bumps illustrating the measures for bump diameter, pitch, and bump height.*

These solder bumps align with corresponding contact pads on the interposer. During reflow, they form electrical connections between the chip and the interposer. The solder bump type is C2, which involves a Copper pillar and a solder cap (Tin). The self-alignment of this process is very poor; however, the electrical connection is present. Fig. 16 shows the SEM image of C2 bumps detailing the intermetallic compound formation. On the left, the solder bump is shown in its entirety, and on the right the connection between the copper pillar and the tin solder is shown. Table 2 presents the materials extracted in the EDS analysis of the C2 micro-bumps.

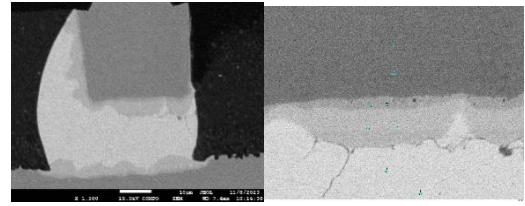

*Figure 16. SEM image of C2 bumps detail showing the intermetallic union.*

*Table 2. EDS analysis results for Cu-Sn mass composition*

| Point | Element | Mass% |
|---|---|---|
| 5 | Cu K | 100 |
| 6 | Cu K | 100 |
| 7 | Cu K | 61.48 |
|   | Sn L | 38.52 |
| 8 | Cu K | 60.37 |
|   | Sn L | 39.63 |
| 9 | Cu K | 36.81 |
|   | Sn L | 63.19 |
| 10 | Cu K | 37.48 |
|    | Sn L | 62.52 |
| 11 | Sn L | 100 |
| 12 | Sn L | 100 |

It is evident that at certain connection points, both materials are present, thus generating an intermetallic compound between tin and copper. Considering the phase diagram of Sn and Cu, the alloys present in this intermetallic compound are determined and will be further discussed in the following section.

The processor's C2 connections are surrounded by epoxy underfill that is dispensed after the chip attachment via solder reflow. The underfill location is marked in Figure 17. Moreover, the whole package is encapsulated using an epoxy over molding compound.

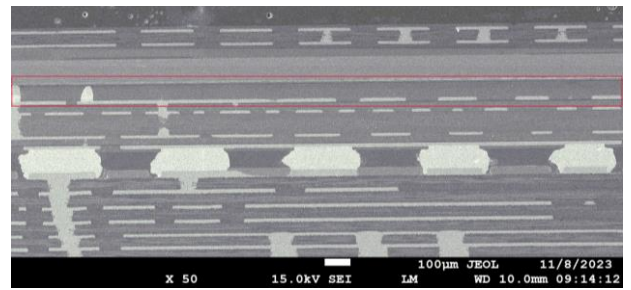

*Figure 17. SEM image of underfill.*

### III. DISCUSSION

During the normal tear-down process, the X-ray analysis would have been done first before the sample preparation for the other analysis. That would have given the ability to define the outline measures of the device as well as the I/O counts for the memory package, the solder connections between the interposers in the processor package as well and the whole device on the PCB level. Even though the total number of these connections stays unknown, the layouts for them could be identified. The schematic illustrations for them are in Fig. 18. The zigzag pattern used with board-level BGA connections is more space-efficient than the common matrix layout and can have more I/Os within the same area. The common matrix in the middle of the same joint level could be for mechanical support, thermal management, and/or for grounding purposes. Also, the extra rows at the ends of the component both for the Memory package's BGA connections and between the processor's interposers are interesting. That could relate to the shape of the device, which is not known for sure, but is assumed to be rectangular instead of square. In that case, the extra rows could be used e.g., for increased reliability and to utilize all the available space to increase the I/O count.

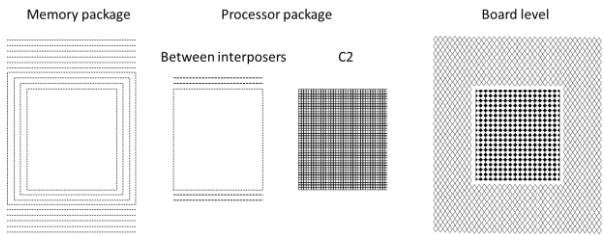

*Figure 18. Layouts for all solder joint groups. From the left: Memory package's BGA connections, C4 connections between the interposers in the processor package, C2 connections for the processor die, and the device level BGA connections.*

The die attaches adhesive layers seen between the stacked memory dies have thin BLTs. For stacking the dies, epoxy adhesives are used due to their hardness, to keep the stack stable. It cannot be identified from the analysis whether the dispensable adhesive or die attach film (DAF) has been used. From the processing point of view using the DAF would be more controlled if the bonding wire layout allows its usage.

The cross-section sample was prepared at an angle (heavily tilted), which is seen from the structures in optical microscope and SEM images, from the sample itself, and in some X-ray images. This makes some analyses, especially some larger measurements, uncertain.

### III.I MEMORY CONNECTIONS

The investigation of the binary phase diagram [13] of Au-Al indicated that the $Au_2Al$ intermetallic compound is formed at wire bonding with this composition. $Au_2Al$ is more ductile and crack resistant which resultantly enhances the reliability of Exynos 990 SoC. However, this intermetallic phase is a poor conductor and thus might result in electrical failure over time.

The reliability of the wire bonds relies on material, mounting process parameters, tools used in mounting, and the structure of the bond [13]. The research shows that a wire bond pad thickness of 5-10µm is better as shown in the Exynos 990 chip [1]. Therefore, the thicker pads provide more material composition for the formation of intermetallic compounds which can reduce the stress concentration and improvement in bond quality.

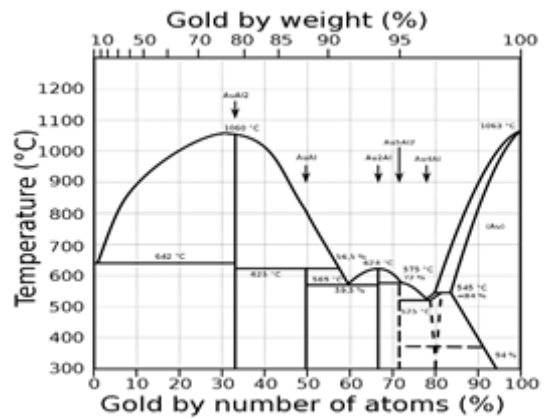

*Figure 19. Binary phase diagram of Au-Al*

Growth of intermetallic Au-Al compounds goes through the consumption of metal layers and results in a reduction in volume and formation of cavities in the contact zone. It has also been reported that the formation of the intermetallic compounds of Au-Al is highly dependent on the contact formation conditions, including temperature, substrate imperfections, and the presence of contaminants [14]. Thus, high cleanliness of the contact area as well as reasonable material thickness are required.

### III.II PROCESSOR CONNECTIONS

The binary phase diagram analysis of Cu-Sn reveals the presence of noteworthy intermetallic compounds at specific points, with $Cu_6Sn_5$ at point 8 and $Cu_3Sn$ at point 7.

The layer closest to the copper pad is a thin layer of $Cu_3Sn$, followed by a thicker layer of $Cu_6Sn_5$ between this and the bulk solder. The $Cu_6Sn_5$ intermetallic compound is particularly significant in soldering Cu-containing base materials with typical Sn-based lead-free solders [15][16]. It has garnered attention due to its advantageous properties, including a high melting point, superior electromigration resistance, thermal stability, and mechanical characteristics [10]. The formation of intermetallic compounds is influenced by various factors, including the size of the bumps. Research indicates that smaller bumps lead to the formation of more intermetallic compounds.

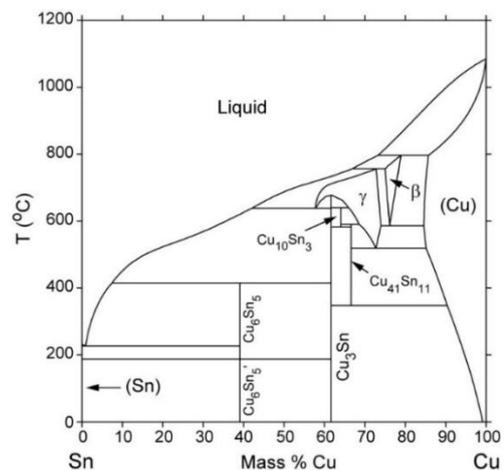

*Figure 20. Binary phase diagram of Sn-Cu*

Samsung uses EMC Epoxy Molding Compound. Delving more into Samsung's EMC it is designed to effectively protect semiconductor circuits from external environmental factors such as moisture, heat, and shock [15]. The EMC is known for its excellent assembly yield improvement, void-less molding in fine pitch bump gap, good gap-filling characteristics with high flowability, and superior releasing performance [17].

In general chip packaging offers high connection density and short interconnection lengths compared to wire bonding, at the cost of more complex manufacturing. Completed chip scale packages are compact, only a couple of millimeters in size. Interconnections are very short to minimize signal delay. Together with the multi-chip design, this FCBGA packaging provides optimum performance density and space efficiency for the Samsung Exynos SoC.

Underfill fills the gap between the die and substrate, surrounding and encapsulating the solder bump interconnects. This helps protect the solder joints from stresses such as fatigue and cracks caused by thermal expansion coefficient mismatch between the die, mold compound, and substrate. The farthest-end flip chip connections face the most stress because of this CTE mismatch [18]. Underfill provides mechanical reinforcement and load sharing to reduce strain on the solder joints during temperature cycling from assembly to board attachment and operation [18].

Compared to metal or ceramic, polymer-based substrates have low stiffness and strength, which is why composites either fillers or particles are used to reinforce the organic substrate material. Not just stiffness, but thermal expansion properties are also altered via this technique [19]. Hence, greater control over the mechanical properties of the organic substrate can be achieved using filler materials.

## IV. Conclusion

This paper reports a teardown analysis of the Samsung Exynos S20 990 SoC, a mobile processor that features a 3D PoP solution with fcPoP. Teardown analysis has contributed to understanding advanced packaging methodologies and indicated potential directions for future semiconductor innovations. For example, typical functionalities of memory, processor and their interconnects have been discussed. Moreover, insights on package type and geometry are also provided to understand the techniques behind the encapsulation of this SoC. We believe that there are still some limitations and challenges that need to be addressed in future research. For example, the reliability and thermal management of the fcPoP design need to be further evaluated through teardown analysis. The compatibility and interoperability of the fcPoP design with other devices and platforms can be tested and verified. The scalability and cost-effectiveness of the fcPoP design need to be optimized and enhanced. These are some of the directions that we suggest for future semiconductor innovations.

# VI. APPENDIX 1

Additional X-ray images with another equipment

9